# Current-Induced Magnetization Switching in MgO Barrier Magnetic Tunnel Junctions with CoFeB Based Synthetic Ferrimagnetic Free Layers

Jun Hayakawa, Shoji Ikeda, Katsuya Miura, Michihiko Yamanouchi, Young Min Lee, Ryutaro Sasaki, Masahiko Ichimura, Kenchi Ito, Takayuki Kawahara, *Member IEEE*, Riichiro Takemura, Toshiyasu Meguro, Fumihiro Matsukura, Hiromasa Takahashi, Hideyuki Matsuoka, Hideo Ohno, *Member IEEE*

*Abstract*—We investigated the effect of using a synthetic ferrimagnetic (SyF) free layer in MgO-based magnetic tunnel junctions (MTJs) on current-induced magnetization switching (CIMS), particularly for application to spin-transfer torque random access memory (SPRAM). The employed SyF free layer had a $Co_{40}Fe_{40}B_{20}$/ Ru/ $Co_{40}Fe_{40}B_{20}$ and $Co_{20}Fe_{60}B_{20}$/ Ru/ $Co_{20}Fe_{60}B_{20}$ structures, and the MTJs ($100 \times (150 - 300)$ nm$^2$) were annealed at 300°C. The use of SyF free layer resulted in low intrinsic critical current density ($J_{c0}$) without degrading the thermal-stability factor ($E/k_B T$, where $E$, $k_B$, and $T$ are the energy potential, the Boltzmann constant, and temperature, respectively). When the two CoFeB layers of a strongly antiferromagnetically coupled SyF free layer had the same thickness, $J_{c0}$ was reduced to $2\text{-}4 \times 10^6$ A/cm$^2$. This low $J_{c0}$ may be due to the decreased effective volume under the large spin accumulation at the CoFeB/Ru. The $E/k_B T$ was over 60, resulting in a retention time of over ten years and suppression of the write current dispersion for SPRAM. The use of the SyF free layer also resulted in a bistable (parallel/antiparallel) magnetization configuration at zero field, enabling the realization of CIMS without the need to apply external fields to compensate for the offset field.

*Index Terms*—current-induced magnetization switching, MgO barrier, CoFeB, synthetic ferrimagnetic free layer

Manuscript received October , 2007. This work was supported in part by the the IT-program of Research Revolution 2002 (RR2002):" Development of Universal Low-Power Spin Memory" from the Ministry of Education, Culture, Sports, Science and Technology of Japan.

J. Hayakawa, K. Miura, M. Yamanouchi, M. Ichimura, K. Ito, H. Takahashi, and H. Matsuoka are with Hitachi, Ltd., Advanced Research Laboratory, Kokubunji 185-8601, Japan (e-mail: jun.hayakawa.pf@hitachi.com)

J. Hayakawa, S. Ikeda, Y. M. Lee, R. Sasaki, T. Meguro, F. Matsukura, and H. Ohno are with the Laboratory for Nanoelectronics and Spintronics, Research Institute of Electrical Communication, Tohoku University, Sendai 980-8577, Japan (e-mail: sikeda@riec.tohoku.ac.jp; ohno@riec.tohoku.ac.jp)

T. Kawahara and R. Takemura are with Hitachi, Ltd., Central Research Laboratory, Kokubunji 185-8601, Japan (e-mail: takayuki.kawahara.rc@hitachi.com)

## I. INTRODUCTION

Spin-polarized currents exert torque on a magnetization that can switch the magnetization direction once the current density becomes sufficiently high.[1],[2] This "current-induced magnetization switching (CIMS)" at reduced current density has been demonstrated in a number of MgO-barrier-based magnetic tunnel junctions (MTJs).[3]-[10] In particular, CoFeB/MgO/CoFeB MTJs have been shown to exhibit high tunnel magnetoresistance (TMR) ratios (over 200%) [11]-[20] together with CIMS at low critical current density ($J_c$). Recently, high-capacity (2-Mb) spin transfer torque random access memory (SPRAM) utilizing the potential of CoFeB/MgO/CoFeB MTJs has been demonstrated [21],[22]. To attain even higher capacity nonvolatile SPRAM, it is necessary to further reduce $J_c$ while maintaining a high thermal-stability factor ($E/k_B T$, where $E$, $k_B$, and $T$ are the energy potential, the Boltzmann constant, and temperature, respectively) well over 60. While $J_c$ is proportional to the product of the magnetization and the thickness of the free layer (i.e., magnetic moment per area), the thermal-stability factor is proportional to the volume of the free layer. [1] If the dimension of an MTJ is simply reduced to enable more bits to accommodate in a given area, $J_c$ remains constant: however, the $E/k_B T$ degrades.

We have been investigating the use of a synthetic ferrimagnetic (SyF) free layer in MgO-barrier-based MTJs to achieve low $J_c$ and high $E/k_B T$. [23],[24] An SyF free layer consisting of two ferromagnetic layers separated by an Ru spacer layer should provide sufficiently high volume to withstand thermal fluctuations while keeping the effective magnetic moment per area low. [25]-[27] Previous studies on CIMS showed that magnetoresistive devices with SyF free layer tends to have a lower $J_c$ than those with a single ferromagnetic free layer. [23],[24],[28] In this paper, we report the CIMS of MgO-barrier-based MTJs with a CoFeB/Ru/CoFeB-based SyF free layer and describe the advantages of the SyF free layer for SPRAM application.

## II. EXPERIMENTAL PROCEDURE

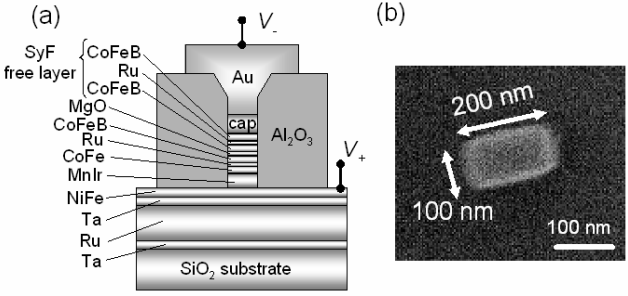
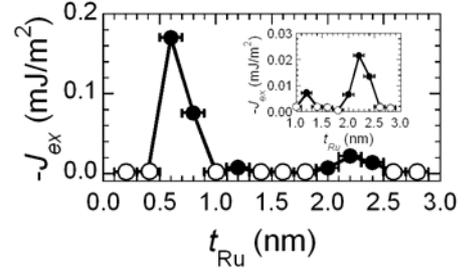

Fig. 1. (a)Schematic drawing of the cross section of the MTJs. The thickness of the Ru spacer layer in the SyF free layer varies from 0.7 to 2.4nm. (b) Scanning electron microscopy image of the pillar.

Fig. 2. Magnetic exchange coupling energy ($J_{ex}$) for $Co_{40}Fe_{40}B_{20}$ /Ru / $Co_{40}Fe_{40}B_{20}$ SyF free layers with Ru spacer layer varying from 0 to 3 nm in thickness. The inset shows expanded view of range from 1.0 to 3.0 nm.

Fig. 1(a) shows a schematic diagram of the MTJ pillar we fabricated for investigating CIMS. MTJ films were deposited on an SiO$_2$/Si substrate by RF magnetron sputtering with a base pressure of $10^{-9}$ Torr. The film layers were, starting from the substrate (in nm), Ta(5)/ Ru(10)/ Ta(5)/ NiFe(5)/ MnIr(8)/ CoFe(2.5)/ Ru(0.8)/ $Co_{40}Fe_{40}B_{20}$(3)/ MgO(0.9)/ $Co_{40}Fe_{40}B_{20}$($t_{CoFeB}$)/ Ru($t_{Ru}$)/ $Co_{40}Fe_{40}B_{20}$($t_{CoFeB}$)/ caping layer. The CoFeB layers and Ru spacer layers were varied in thickness from 1.8 to 2.6 nm and 0.6 to 2.4 nm, respectively, and were formed by using a slide mask shutter during sputtering, respectively. The nano-scaled junctions (100 x 150 nm$^2$, 100 × 200 nm$^2$, and 100 × 300 nm$^2$) were fabricated using electron-beam lithography. Fig. 1(b) shows a scanning electron microscopy image of a rectangular MTJ pillar 100 × 200 nm$^2$. The completed MTJs were annealed at 300$^o$C for 1 h in a $10^{-6}$ Torr vacuum under a magnetic field of 4 kOe. The TMR loops of the MTJs were measured at room temperature using a four-probe method with dc bias and a magnetic field of up to 1 kOe. The CIMS was evaluated by measuring the resistance by 10-μA-step current pulses with a duration ($\tau_p$) ranging from 30 μs to 1 s. The $E/k_BT$ was obtained from the slope of the average critical current density ($J_c^{ave}$) vs. ln($\tau_p/\tau_0$) plot and from $J_c^{ave}$ vs $H_c$ plot, where $J_c^{ave}$ is defined as ($|J_c^{P->AP}|+|J_c^{AP->P}|$)/2. The current direction was defined as positive when the electrons flowed from the top (free) to the bottom (pin) layer.

### III. EXPERIMENTAL RESULTS AND DISCUSSION

To determine the magnetic exchange coupling energy ($J_{ex}$) between the two CoFeB ferromagnetic layers as a function of $t_{Ru}$ in the CoFeB(2)/Ru($t_{Ru}$)/CoFeB(2) structure, we prepared separately a structure, SiO$_2$/Si substrate / Ta(5)/ Ru(50)/ Ta(5)/ MgO(0.9)/ CoFeB(2)/ Ru($t_{Ru}$)/ CoFeB(2)/ Ta(5). We cut it into a rectangle (1 × 3 mm$^2$) for magneto-optical Kerr effect measurements. Fig. 2 plots the $J_{ex}$ between the two layers as a function of $t_{Ru}$. The inset shows an expanded view for $t_{Ru}$ = 1.0 to 3.0 nm. We calculated $J_{ex}$ using $J_{ex}$ = –$\mu_0 H_s M_s t_1 M_2 t_2/(M_1 t_1+M_2 t_2)$, where $H_s$ is the saturation field, $M_1$ and $M_2$ are the saturation magnetizations of the two CoFeB (1.3 T), $t_1$ and $t_2$ are the thicknesses of the two layers (both 2 nm). [4],[29] The highest antiferromagnetic coupling energy (0.17 mJ/m$^2$) was obtained for $t_{Ru}$ ~0.6 nm. There were oscillations in the magnitude of $J_{ex}$; a second peak appeared at $t_{Ru}$ = 1.2 nm and a third one at 2.4 nm. The oscillatory behavior suggests the presence of an oscillation from ferromagnetic to antiferromagnetic and back, as previously reported [30]. It originates from the Ruderman-Kittel-Kasuya-Yosida (RKKY)-type coupling typically found in Co/Ru/Co multilayers. [31] The open circles in Fig. 2 represent either no $J_{ex}$ or positive (ferromagnetic) $J_{ex}$ because we cannot measure the ferromagnetic coupling with the method described here.

Fig. 3 plots $J_c^{ave}$ as a function of 1/$H_c$ (inverse of coercivity, $H_c$) for all the MTJs shown in Fig. 2, with Ru spacers from 0.6 to 2.4 nm. The $J_c^{ave}$ was measured with a pulse current of 1 s duration. $H_c$, which were obtained from TMR measurement under magnetic field, are varied by changing the $t_{Ru}$ and aspect ratio of MTJ pillars. We see that $J_c^{ave}$ can be categorized into three groups depending on the strength of the magnetic coupling energy shown in Fig. 2 and that, within each group, $J_c^{ave}$ increased linearly with 1/$H_c$. The black symbols ($t_{Ru}$ = 0.7 and 0.9 nm), white symbols ($t_{Ru}$ = 1.5, 1.7, and 1.9 nm), and hatched square symbols ($t_{Ru}$ = 2.2 and 2.4 nm) correspond to the Ru spacer thicknesses for the first antiferromagnetic coupling, the ferromagnetic couplings between the second and third antiferromagnetic coupling, and the third antiferromagnetic coupling, respectively. Hereafter, we refer to the groups as Group I, II, and III, respectively.

We now show that the intrinsic critical current density ($J_{c0}$) can be obtained by analyzing the measured $J_c^{ave}$ vs. $H_c$ plot using Slonczewski's model [1],[32] taking into account the thermal activated nature of the magnetization switching. The relevant equations are [33],[34]

$$J_c = J_{c0}\{1-(k_BT/E)\ln(\tau_p/\tau_0)\}, \quad (1)$$

$$J_{c0} = \alpha\gamma eM_s t(H_{ext} \pm H_k \pm H_d)/\mu_B g, \quad (2)$$

$$E = M_sVH_k/2, \quad (3)$$

$$g = P/[2(1+P^2\cos\theta)], \quad (4)$$

where $\alpha$ is the Gilbert damping coefficient, $\gamma$ the gyromagnetic constant, $e$ the elementary charge, $t$ the thickness of the free layer, $H_{ext}$ the external magnetic field, $H_k$ the in-plane uniaxial magnetic anisotropy, $M_s$ the saturation magnetization of the free layer, $V$ the volume of the free layer, and $H_d$ the out-of-plane magnetic anisotropy induced by the demagnetization field. The $\theta$ is 0 for the parallel configuration and $\pi$ for anti-parallel.

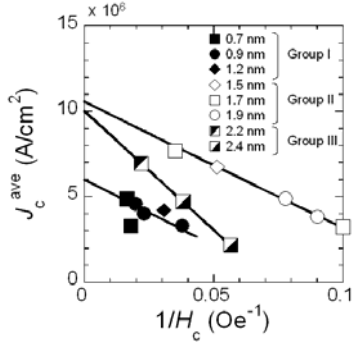

Fig. 3. $J_c^{ave}$ as a function of $1/H_c$ The plotted black symbols ($t_{Ru}$ = 0.7, 0.9, and 1.2nm), white symbols ($t_{Ru}$ = 1.5, 1.7, and 1.9 nm), and hatched square symbols ($t_{Ru}$ = 2.2 and 2.4 nm).

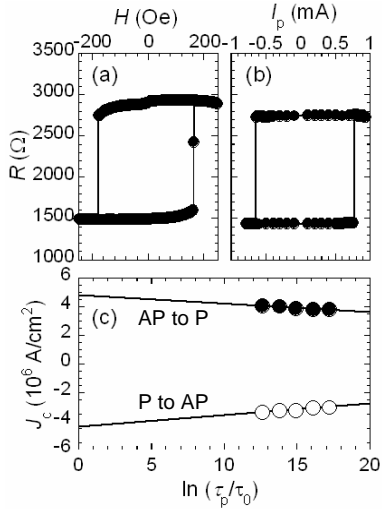

Fig. 4. (a) $R$-$H$ loop, (b) $R$-$I_p$ loop at $\tau_p$ of 1 ms, and (c) $J_c$ as a function of ln($\tau_p/\tau_0$) for an MTJ with $Co_{40}Fe_{40}B_{20}(2)/Ru(0.8)/Co_{40}Fe_{40}B_{20}(2)$ SyF free layer.

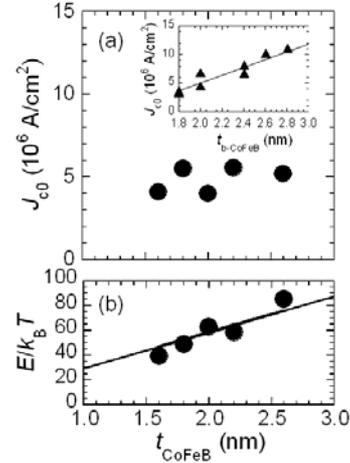

Fig. 5. (a) $J_{c0}$ and (b) $E/k_B T$ as a function of $t_{CoFeB}$ for an MTJ with $Co_{40}Fe_{40}B_{20}(t_{CoFeB})/Ru(0.8)/Co_{40}Fe_{40}B_{20}(t_{CoFeB})$ SyF free layer. The inset of (a) shows $J_{c0}$ as a function of the bottom-CoFeB thickness when the thickness of top CoFeB is fixed at 1.8 nm.

In the following, we deal with a magnetic cell with $E/k_B T$ enough high to have a uni-axial anisotropy and a single magnetic domain, so $H_k \approx H_c$. Since $H_d \gg H_k$, $H_{ext}$, we can obtain the $J_{c0}^{ave.}$ and $J_c^{ave}$, which is a function of $H_c$, using

$$J_{c0}^{ave} \approx \alpha \gamma e M_s t H_d (g^{(P \to AP)} + g^{(AP \to P)}) / 2\mu_B g^{(P \to AP)} g^{(AP \to P)} \quad (5)$$

$$J_c^{ave} = J_{c0}^{ave} [1 - (2k_B T/M_s V) \ln(\tau_p/\tau_0)/H_c] \quad (6)$$

As seen in Fig. 3, the measured $J_c^{ave}$ is inversely proportional to $H_c$ within each group (solid lines). This shows that eq. (6) is a good approximation of our results. By extrapolating $1/H_c$ to zero in Fig. 3, we can obtain $J_{c0}^{ave}$, given by eq. (5), as $6.0 \times 10^6$ A/cm$^2$ for Group I, $1.00 \times 10^7$ A/cm$^2$ for Group II, and $1.06 \times 10^7$ A/cm$^2$ for Group III. The $J_{c0}$ for Group I was the smallest indicating that the larger the antiferromagnetic coupling, the lower the $J_{c0}$. We speculate that the lower $J_{c0}$ for Group I may be responsible for both the lower effective product of magnetization and thickness (volume) and the larger spin accumulation at CoFeB-Ru interface. Two antiferromagnetically coupled CoFeB layers separated by a non-magnetic Ru layer much thinner than the spin diffusion length [35] is known to enhance the spin accumulation at the CoFeB-Ru interface.[36],[37] Spin accumulation increases the efficiency of the spin-torque acting on the CoFeB free layer and contributes to the reduction in critical current density. The effect of the effective product of magnetization and thickness on $J_{c0}$ will be discussed later.

Figs. 4 (a) and (b) show the magnetic field hysteresis loop ($R$-$H$ loop) and the resistance vs. pulsed current ($R$-$I_p$) for $\tau_p$ = 1 ms and a 100 × 200 nm$^2$ MTJ with a CoFeB(2)/Ru(0.8)/CoFeB(2) SyF free layer. The Ru thickness of 0.8 nm corresponds to the highest antiferromagnetic coupling energy. The TMR ratio was 98%. The $R$-$I_p$ curves were measured under an applied magnetic field of –7 Oe in the direction of the pin CoFeB layer to compensate for the offset field [see Fig. 4(a)] arising primarily from the stray fields at the edge of the patterned SyF pin layer. The current density required to switch the magnetization from parallel (anti-parallel) to anti-parallel (parallel) shown in Fig. 4(b) was $3.85 \times 10^6$ A/cm$^2$ ($J_c^{AP \to P}$ = –3.25 x 10$^6$ A/cm$^2$); the $J_c^{ave}$ was $3.55 \times 10^6$ A/cm$^2$. Fig. 4(c) plots $J_c^{ave}$ as a function of ln($\tau_p/\tau_0$) for $\tau_p$ from 100 μs to 1 s. The slope of the lines in Fig. 4(c), which is based on eq. (1), indicates an average $E/k_B T$ 68. By extrapolating $J_c$ to an ln($\tau_p/\tau_0$) of 0, which corresponds to $\tau_p$ = 1 ns, we obtain an $J_{c0}^{ave}$ of 4.6 x 10$^6$ A/cm$^2$. The $J_{c0}^{ave}$ thereby obtained agrees well with that obtained from Fig. 3.

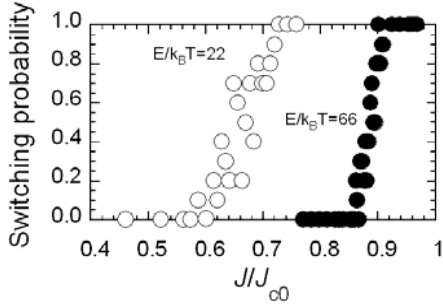

Fig. 6. Switching probability as a function of $J/J_{c0}$ for CoFeB/Ru/CoFeB SyF free layer-based MTJs with different $E/k_BT$ values. The applied current pulse duration is 1 μs.

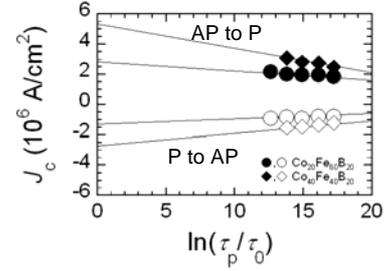

Fig. 8. $J_c$ as a function of $\ln(\tau_p/\tau_0)$ for MTJs with $Co_{20}Fe_{60}B_{20}$/ Ru/ $Co_{20}Fe_{60}B_{20}$ and $Co_{40}Fe_{40}B_{20}$/ Ru/ $Co_{40}Fe_{40}B_{20}$ SyF free layers.

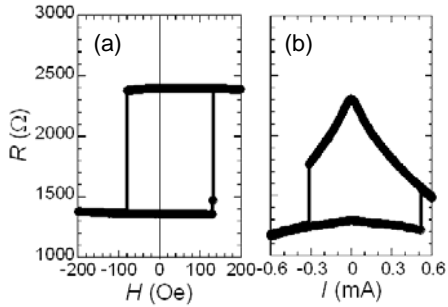

Fig. 7. (a) $R$-$H$ loop and (b) $R$-$I$ loop measured at zero magnetic field for MTJs with $Co_{40}Fe_{40}B_{20}(1.8)$/ Ru(0.8)/ $Co_{40}Fe_{40}B_{20}(1.8)$ SyF free layer.

Table. I  Comparison of $J_{c0}/\{(E/k_BT) \times A\}$

| Structure of free layer | $J_{c0}/(E/k_BT)/A$ ($10^{15}$ A) | |
|---|---|---|
| CoFeB/Ru/CoFeB SyF | 0.70 | |
| Single CoFeB | 0.85 | ref.[8] |
|  | 1.7 | ref.[21] |
|  | 4.8 | ref.[6] |

We now discuss the effect of the CoFeB ferromagnetic layer thickness ($t_{CoFeB}$) in the SyF free layer on $J_{c0}$ and $E/k_BT$. Fig. 5 shows $J_{c0}$ and $E/k_BT$ as a function of $t_{CoFeB}$ in MTJs with a CoFeB($t_{CoFeB}$)/ Ru(0.8)/ CoFeB($t_{CoFeB}$) SyF free layer, in which the two CoFeB layers have nominally the same thickness. MTJs with dimensions of 100 × 200 nm² were selected so as to have almost the same $H_c$ (~150 Oe). The inset of Fig. 5 (a) shows $J_{c0}$ vs. bottom CoFeB thickness when the top CoFeB thickness was fixed at 1.8 nm. Fig. 5 (a) shows that the $J_{c0}$ for all samples was apparently constant with regards to $t_{CoFeB}$. In contrast, it increased linearly with an increase in the bottom CoFeB thickness, as shown in the inset. These results suggest that $J_{c0}$ is proportional to the effective product of magnetization and thickness ($Mt_{eff}$) of the SyF free layers. The effective thickness ($t_{eff}$) is given by $t_{eff} = |Mt_{t-CoFeB} - Mt_{b-CoFeB}| / M_{eff}^{SyF}$, where subscripts t and b refer to the top and bottom CoFeB layers of the SyF free layer. [38] In contrast, $E/k_BT$ increased linearly with an increase in $t_{CoFeB}$ as shown in Fig. 5 (b), indicating that $E/k_BT$ in the SyF free layer is determined by the total CoFeB thickness of the free layer. We obtained a high $E/k_BT$ (over 80) at $t_{CoFeB}$ = 2.6 nm, which is high enough to endure a retention time of over ten years. The SyF free layer enables the realization of a high $E/k_BT$ by increasing $t_{CoFeB}$ without increasing $J_{c0}$.

For SPRAM, the suppression of the write current dispersion is one of the most important issues. The magnetization is potentially switched by a current during data reading, because the magnetization switching of the free layer occurs stochastically in the thermal-activation region of reading and writing times of over 10 ns. From eq. (7) [39], a high $E/k_BT$ of SyF free layer should lead to suppress the dispersion of the write current.

$$P = 1 - \exp\{-(\tau_p/\tau_0)\exp[-E/k_BT(1-J/J_{c0})]\} \quad , \quad (7)$$

where $P$ is the switching probability. Fig. 6 shows $P$ under $\tau_p$ = 1 μs as a function of $J/J_{c0}$ for MTJs with $Co_{40}Fe_{40}B_{20}$/ Ru/ $Co_{40}Fe_{40}B_{20}$ SyF free layer with different $E/k_BT$ (22 and 66). The slope of each curve becomes steeper with increasing $E/k_BT$. This indicates that the dispersion of write current is further suppressed by employing the SyF free layer with even higher $E/k_BT$ in MTJs.

Another remarkable advantage of the SyF free layer for SPRAM is described here. For the CIMS of the MTJs, an external magnetic field is usually needed to compensate for the shift of the zero magnetic field generated by interlayer coupling between the free and pinned layer and by the stray file from the pinned layer. Especially for MTJs with a single free layer, a bistable (parallel/antiparallel) magnetic configuration is hard to achieve due to the large shift of the zero magnetic field and the small $H_c$. Fig. 7 (a) shows a typical $R$-$H$ loop for an MTJ with a CoFeB(1.8)/ Ru(0.8)/

CoFeB(1.8) SyF free layer. It shows that a bistable magnetization configuration at zero field can be stably formed. This behavior comes from the closed magnetic field within an SyF free layer and the large $H_c$ (typically over 100 Oe). Fig. 7 (b) shows *R-I* loop measured under a zero field. Clear switching is evident. Thus, the SyF free layer enables the realization of CIMS without the needed to apply external fields to compensate for the offset field.

Finally, we also examined the CIMS in MTJs by replacing the $Co_{40}Fe_{40}B_{20}$ layer in the SyF free layer with an Fe-rich $Co_{20}Fe_{60}B_{20}$ layer. TMR ratio (130%) of MgO-barrier MTJs with $Co_{20}Fe_{60}B_{20}$ electrode is higher than that (~90%) of MTJs with $Co_{40}Fe_{40}B_{20}$ electrode, suggesting that MTJs with $Co_{20}Fe_{60}B_{20}$ electrode should have a lower $J_{c0}$ [20], [32]. Fig. 8 shows the $J_c$ as a function of for MTJs ($100 \times 200$ nm$^2$) with two kinds of SyF free layers ($Co_{20}Fe_{60}B_{20}(2)$/ Ru(0.8)/ $Co_{20}Fe_{60}B_{20}(1.8)$ (●) and $Co_{40}Fe_{40}B_{20}(2)$/ Ru(0.8)/ $Co_{40}Fe_{40}B_{20}(1.8)$ (♦). From the slopes, the obtained $J_{c0}$ ($E/k_BT$) are $2.0 \times 10^6$ A/cm$^2$ (47) for MTJ with $Co_{20}Fe_{60}B_{20}$ SyF free layer and $4.1 \times 10^6$ A/cm$^2$ (42) for $Co_{40}Fe_{40}B_{20}$. Not only a higher TMR ratio but also a lower effective magnetization ($M_{eff}$) and a lower damping factor ($\alpha$) in $Co_{20}Fe_{40}B_{20}$ compared to those in $Co_{40}Fe_{40}B_{20}$ may be responsible to further decrease in $J_{c0}$.

Here, we suggest the use of parameter $J_{c0}/\{(E/k_BT) \times A\}$ (where *A* is junction area) as a figure of merit for in the comparison of different MTJs, because one needs to achieve both low $J_{c0}$ and high $E/k_BT$ per area *A*. Table I shows $J_{c0}/\{(E/k_BT) \times A\}$ obtained in the present study as well as in previous studies on CIMS. The smaller value indicates better performance. The value obtained in the MTJs with SyF free layer layer is the smallest among all.

## III. CONCLUSION

We have described the advantages of MgO-barrier-based MTJs with a $Co_{40}Fe_{40}B_{20}/Ru/Co_{40}Fe_{40}B_{20}$ SyF free layer in terms of current-induced magnetization switching (CIMS) and their application to spin-transfer torque random access memory (SPRAM). One advantage is low intrinsic critical current density ($J_{c0}$) without degrading thermal-stability factor ($E/k_BT$). When the two CoFeB layers in a strongly antiferromagnetically coupled SyF free layer have the same thickness, a low $J_{c0}$ of $4 \times 10^6$ A/cm$^2$ is observed. Replacing the $Co_{40}Fe_{40}B_{20}$ electrode with $Co_{20}Fe_{60}B_{20}$ one further reduces it to $2 \times 10^6$ A/cm$^2$. We believe that this low $J_{c0}$ is caused by the decreased effective volume under the large spin accumulation at the CoFeB/Ru interface. The high $E/k_BT$ (over 60) obtained in the SyF free layer results in a retention time of over ten years and suppression of the write current dispersion for SPRAM. Another remarkable advantage of the SyF free layer is that it results in a bistable (parallel/antiparallel) magnetization configuration at zero field, making it possible to realize CIMS without the need to apply external fields to compensate for the offset field.

ACNOWLEDGEMENT

The authors would like to thank Dr. Yuzo Ohno and Dr. Takehiro Tanikawa for valuable helps with the experiment and discussion.